\documentclass[]{tCPH2e}

\begin{document}

\title{New nuclear magic numbers}

\author{Reiner Kr\"ucken$^{\ast}$\thanks{$^\ast$Email:
reiner.kruecken@ph.tum.de\vspace{6pt}}\\{\em{Physik Department E12,
Technische Universit\"at M\"unchen, Garching,
Germany}}\\\vspace{6pt}\received{May 2010} }

\maketitle

\begin{abstract}
The nuclear shell model is a benchmark for the description of the structure of atomic nuclei.
The magic numbers associated with closed shells have long been assumed to be valid across the whole nuclear chart.
Investigations in recent years of nuclei far away from nuclear stability at facilities for radioactive ion beams have
revealed that the magic numbers may change locally in those exotic nuclei leading to the disappearance of classic
shell gaps and the appearance of new magic numbers. These changes in shell structure also have important implications
for the synthesis of heavy elements in stars and stellar explosions. In this review a brief overview of the basics of the nuclear
shell model will be given together with a summary of recent theoretical and experimental activities investigating
these changes in the nuclear shell structure.
\bigskip
\bigskip

\end{abstract}

\section{Introduction}
The atomic nucleus, core of the atom, is the carrier of essentially
all visible mass in the Universe. Atomic nuclei are also the fuel
for the burning of all stars and are thus the source of the energy
enabling live on this planet. Nuclear reactions are also at the
heart of the production of all naturally occurring chemical elements
from deuterium to uranium.

Despite this central role of atomic nuclei and nuclear reactions in
the Universe, the structure and dynamics of atomic nuclei cannot yet
be satisfactorily described on the basis of the fundamental
underlying theory of the strong interaction, quantum chromodynamics
(QCD). This results from the fact that the protons and neutrons
inside the atomic nucleus are themselves complex systems built up
from quarks and gluons, the force carriers of the strong
interaction. Thus the strong interaction between nucleons is an
effective, van der Waals like interaction that governs a
two-component quantum many-body system, making analytical
calculations impossible. The free nucleon-nucleon interaction can be
measured via scattering experiments and is well described
theoretically. However, inside an atomic nucleus the interaction
between two nucleons changes due to the presence of the other
nucleons or, in other word, the nuclear matter in which the nucleons
are submerged. Therefore, effective interactions have to be used to
describe the interaction of protons and neutrons inside an atomic
nucleus. Much progress has been made in recent years to derive
effective interactions that are motivated by the symmetry properties
of QCD but which also take into account the effects of the nuclear
medium on the nucleon-nucleon interaction
\cite{Bed02,Epe06,Fin06,Epe09}. It is one of the central goals of
modern nuclear physics to develop a unified theoretical framework
that allows one to reliably predict the properties of complex nuclei
with a set of theoretical tools, which are based on the same
underlying basic ingredients connected to the QCD symmetry properties.

In the recent decades nuclear structure physics has undergone a major
re-orientation and rejuvenation seeing the discovery of new
phenomena and the emergence of new frontiers. In particular the
availability of energetic beams of short-lived (radioactive) nuclei,
in the following also referred to as rare isotope or exotic nuclear
beams, has opened the way for the exploration of the structure and
dynamics of complex nuclei in regions far away from stability, where
very limited information is available. These new experimental
capabilities allow one to delineate the limits of nuclear existence and
study the dependence of the nuclear force on proton and neutron
number (or isospin\footnote{Proton and neutron are treated as two
different projections $T_z= -1/2$ and $T_z=+1/2$, respectively, of
an isospin $T=1$ state. }). They also enable us to study the
behavior of nuclei near and beyond the neutron and proton drip-lines
and to investigate the emergence of new modes of nuclear behavior,
not observed near stability. The neutron (proton) drip-line denotes the line beyond which
the separation energy $S_n$  ($S_p$) for the last neutron (proton)
becomes negative and thus no additional neutron (proton) can be bound
by the nuclear force \footnote{Note that a proton with $S_p < 0$ can
be bound to the nucleus by the Coulomb barrier. Such nuclei can decay
by proton radioactivity, the tunneling of the proton through the Coulomb barrier \cite{Woo97}.}.
Examples for such new modes are nuclear
halos, skins of protons or neutrons, and new modes of excitation
were protons and neutrons decouple as well as new isospin pairing
phases and new decay modes
\cite{Han03,Aum05,Kee07,Paa07,Ben07,Bla08,Gad08,Kee09}.

\begin{figure}
\begin{center}
\includegraphics[width=\textwidth]{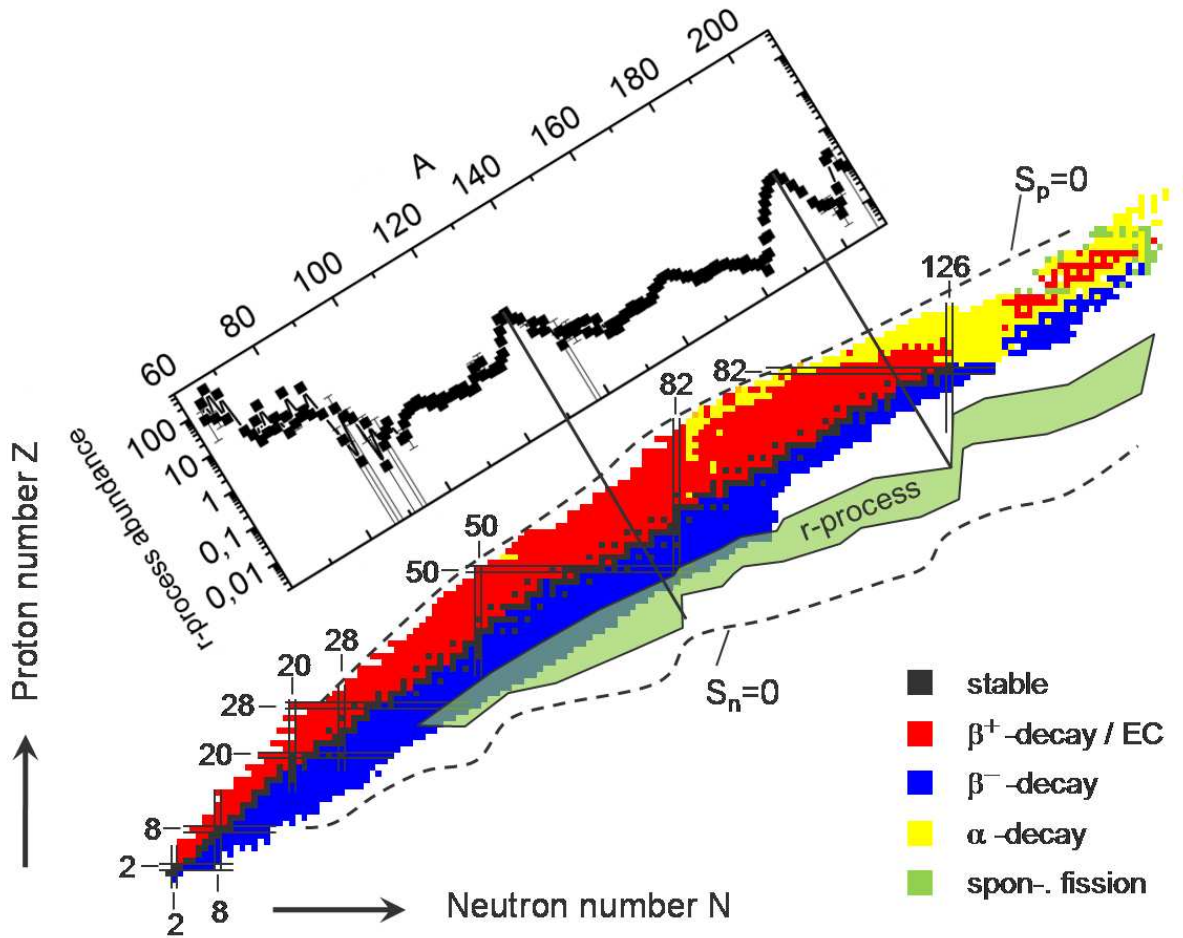}
\caption{Nuclear chart with indicated classic shell closures and r-process path.
The dashed lines indicate roughly where we believe that nuclei are bound or in
other words that the neutron and proton separation energies $S_n$, $S_p$ are positive.
Also shown are solar r-process abundances, taken from \protect\cite{Kae89}.}%
\label{fig:chart}
\end{center}
\end{figure}
%FIG nuclear chart with shells indicated.
%(needs to be discussed together with central questions (limits of stability,

One of the major successes in the description of the properties of
atomic nuclei is the introduction of the nuclear shell model. The
shell structure was observed (see e.g. \cite{Els34,Goe48}) in the
occurrence of special numbers of protons and neutrons in nuclei, so
called magic numbers, for which discontinuities appear, e.g. in
binding energies and separation energies (the energies needed to
remove the last neutron or proton). Also other experimental
observables, such as energies of first excited states in even-even nuclei, and
electromagnetic transition rates show these shell effects. By
analogy to the very successful atomic shell model these magic
numbers were taken as evidence of closed shell configurations in
atomic nuclei, similar to the noble gas configurations in the electron shells of atoms. The
classical magic numbers are indicated as horizontal and vertical
lines in the nuclear chart shown in Fig. \ref{fig:chart}.

Until recently the persistence of these magic numbers throughout the
whole nuclear chart was a fundamental paradigm in nuclear physics.
However, in the last two decades sophisticated experiments with
exotic nuclear beams made it possible to scrutinize the persistence
of the shell gaps far away from stability. The measurements showed
some surprising changes in the nuclear shell structure as a function
of proton and neutron number in light nuclei. These observations
triggered numerous theoretical investigations, which in turn made
new predictions that some magic numbers will disappear and new shell
gaps will appear in certain regions of the nuclear chart.

The shell structure in nuclei plays also an important role in
defining the pathways on the nuclear chart of various processes of the
synthesis of the chemical elements in the Universe (see \cite{Gra07}
for a recent review) and results in specific features of the
observed solar abundances \cite{BBFH}.

In particular, the possibility of a quenching of shell gaps in
neutron-rich medium mass and heavy nuclei has been discussed in the
context of reproducing the r-process abundances
\cite{Che95,Pfe97,Pfe01} in network calculations. The rapid neutron
capture (r-)process is responsible for the production of about half
of the heavy elements above iron. It proceeds in a very
neutron-rich, high entropy environment where certain seed nuclei
capture neutrons until an equilibrium abundance of isotopes for an
element (Z) is reached between neutron capture and
photo-dissociation in the hot environment \cite{Cow91}. Beta-decay
out of this equilibrium distribution leads to the next heavier
element (Z+1) where again neutron capture occurs until an
equilibrium abundance distribution is reached and beta decay leads
to the next heavier element. By this mechanism all heavy elements
above iron can be produced. The most abundant isotope in an isotopic
chain under equilibrium conditions can be estimated for a given
neutron flux by a characteristic value of the neutron separation
energy $S_n$ \cite{Gra07} with the equation
$$
S_n (\mbox{MeV}) = \frac{T_9}{5.04}\left( 34.075 - \log n_n + \frac{3}{2}\log T_9 \right),
$$
where $T_9$ is the temperature in units of $10^9$ K and $n_n$ is the
neutron density in cm$^{-3}$. This equation implies that the
r-process proceeds along a path of constant neutron separation
energies, which is indicated in Fig. \ref{fig:chart}. Once the
neutron density drops sufficiently the process falls out of
equilibrium. After the neutron capture stops for lack of neutrons
the neutron-rich nuclei decay via beta decay towards stability to
produce the observed r-process abundances.

The shell structure along the r-process path is imprinted in the
abundance pattern, since the beta-decay half-lives drop
significantly just after the shell closure and thus abundance is
collected along the neutron-shell closure. The broad abundance peaks
in the r-process abundances at masses A=130 and A=195 (see Fig.
\ref{fig:chart}) are a result of this. If, for example, the shell
closures at N=82 and N=126 persist in neutron-rich nuclei, the
r-process path will flow along these shell closures at these mass
numbers. However, if the shell gaps are quenched or if shell gaps
occur for different neutron (or proton) numbers, the path of
constant separation energies may be shifted and the r-process path
may be located further away or closer to stability.

Thus, the underlying nuclear physics of nuclei far off stability can
have a significant impact on pathways of nucleosynthesis. Certainly
the astrophysical conditions at the nucleosynthesis sites will be
decisive for the element production. In case of the r-process the
astrophysical site has not yet been clearly identified. One of the
favorite scenarios is the neutrino-driven wind in core-collapse
supernovae of massive stars \cite{Mey92,Woo92,Pan08}. However,
neutron-star mergers are also being considered as a possible site
\cite{Sur08}. It will take the combined effort of multidimensional
astrophysical modeling, precision astronomical observations of
elemental and isotopic abundances as well as precise knowledge of
the nuclear physics of exotic nuclei to solve the puzzle of the
origin of the heavy elements.

In this article I will try to provide a brief review of the recent
developments concerning the modification of shell structure far away
from stability. I will start by recapitulating some of the basic
features of the nuclear shell model and by introducing the main
ingredients for shell model calculations. On the basis of these
foundations I will try to summarize how shell structure may change
as one goes away from stability. After that I will provide some
overview of how one may look experimentally for shell closures and
provide examples of recent experiments that have helped to discover
the disappearance of classic shells and the appearance of new magic
numbers.

\section{Basics of the nuclear shell model}

The underlying concept for the nuclear shell model is the idea that
nucleons move almost independently inside a central potential well.
However, contrary to atomic physics, where the central potential for
the electrons originates from the charge of the atomic nucleus, the
central potential for a nucleon is the result of its interactions
with the other nucleons in the atomic nucleus.

\subsection{Independent particle model}

\begin{figure}
\begin{center}
\includegraphics[width=85mm]{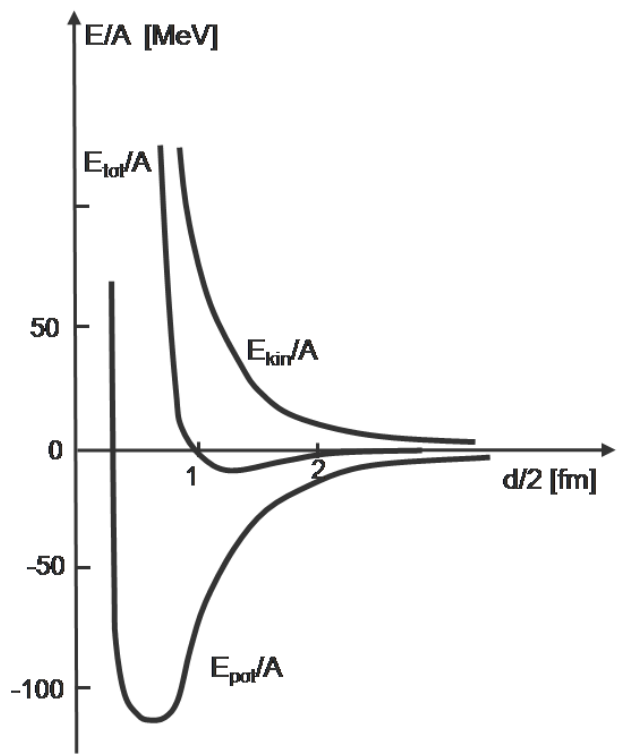}
\caption{Potential and kinetic part of the radial part of the nucleon-nucleon interaction
resulting in a total energy with a weak binding at distances of 1.5-2 fm
(Adopted from \protect\cite{Rin80} ).}%
\label{fig:NNint}
\end{center}
\end{figure}

\begin{figure}
\begin{center}
\includegraphics[width=85mm]{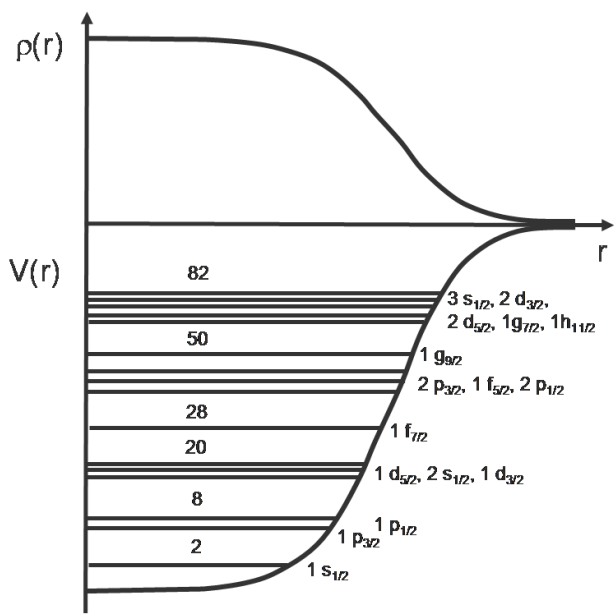}
\caption{Woods-Saxon type mean field with clustering of single-particle levels
into major shells. In the top of the figure the corresponding radial density distribution
is indicated.}%
\label{fig:mean-field}
\end{center}
\end{figure}

The concept of independent particle motion seems surprising if one
looks at the well known radial dependence of the
nucleon-nucleon potential (see fig. \ref{fig:NNint}) with its short
range repulsion (at about 0.5 fm) and its strongly attractive
component, which can be described by pion exchange. Aside from this
central part the nucleon-nucleon interaction also contains
spin-orbit, spin-spin, and tensor components\footnote{ The tensor
interaction will play an important role later on in this review.}.
As a result of the strong short range repulsion the nucleons have a
large kinetic energy at short distances and the total energy, as sum
of potential and kinetic energy, shows only a relatively weak
binding at distances of 1.5-2 fm. This saturation property of the
nuclear force also leads to the almost constant density of the
nuclear matter inside nuclei. The resulting central potential can be
approximately described by an inverted Fermi-function, the so-called
Woods-Saxon potential (WS) (see Fig. \ref{fig:mean-field}) or
alternatively by a harmonic oscillator (HO) potential with an
additional term proportional to the square of the orbital angular
momentum, in order to effectively flatten the potential in the
nuclear interior.

% Fig with NN-potential + kinetic term + total energy (a la Ring)

As a result of the central potential one obtains groups of energy
levels forming shells separated by large energy gaps as indicated
in Fig. \ref{fig:mean-field}. By filling these
individual shells consistent with the requirements of the Pauli
exclusion principle one obtains closed inert configurations, similar
to the noble gas configurations in atoms, for certain numbers of
protons and neutrons, the so called magic numbers (2, 8, 20, 28, 50,
82, 126). The first attempts to explain the observed magic numbers
with a nuclear shell model failed until in 1949 Mayer, Haxel, Suess
and Jensen \cite{Goe49,Hax49} showed that the inclusion of a strong
spin-orbit interaction with a sign opposite to that known from the
fine structure in atoms gave rise to the observed gaps between the
nuclear shells. The original form of this spherical mean field was:
\begin{equation}
U(r) = \frac{1}{2} m \omega^2 r^2 + D\vec{\ell}^2+C\vec{\ell}\cdot\vec{s}.
\end{equation}

% Fig with Woods-Saxon Potential and nuclear density distribution

The exact ordering within the shells depends on the choice of
potential, HO vs. WS, and their parameters. Please note also that
there is a mass dependence in these parameters, most obvious due to
the change of the radius of the potential well proportional to
$A^{1/3}$. Today a more realistic mean field potential can be
self-consistently calculated using Hartee-Fock methods based on
effective nucleon-nucleon interactions \cite{Ben03,Vre05}.

Already with the concept of independent particles moving in this
potential, a one-body mean field potential, a broad body of
experimental data can be explained, in particular for nuclei with
only one particle or hole outside a magic core, e.g. ground state
spins, excited states, magnetic moments. However, this model
certainly is too naive since the potential well cannot account for
all aspects of the interaction between the nucleons. This leads to
so-called core-polarization effects where the
particles outside the magic core will have some influence on the
core itself, which can have dramatic effects, e.g. on magnetic
moments. In addition one has to take into account a residual
interaction among the nucleons outside the magic core. This residual
interaction plays a major role in the excitation spectrum of nuclei
away from the magic shells and will be discussed in the following
subsection.

\subsection{Interacting shell model}

\begin{figure}
\begin{center}
\includegraphics[width=85mm]{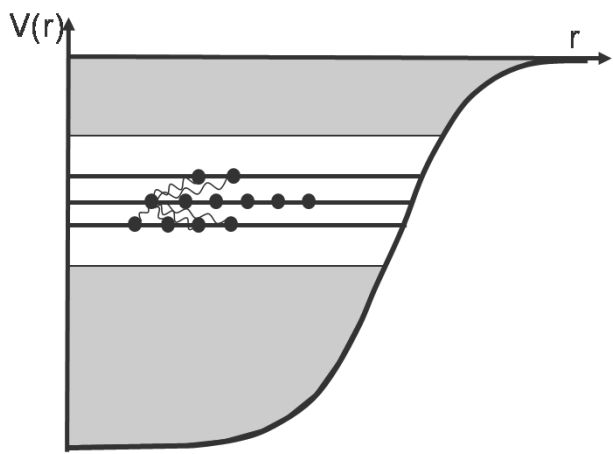}
\caption{Basic ingredient for an interacting SM calculation.
Single-particle energies (SPE) in the configuration space and two-body-matrix elements (TBME)
of the interaction between the nucleons in the configuration space. A closed shell core and
unoccupied orbits above the configuration space are not part of the SM diagonalization and are colored in grey. }%
\label{fig:interact-SM}
\end{center}
\end{figure}

In a realistic shell model calculation the residual interaction has
to be taken into account between all valence particles outside an
appropriately chosen closed core of fully occupied shells. For this
appropriate choice one has to ensure that excitations from this
closed configuration do not play a role in the excitation spectrum
of the nucleus. In order to perform an interacting shell model
calculation one defines a configuration space of single-particle
orbits for the valence nucleons, which can interact. The
configuration space is limited in size by taking into account only a
limited number of orbits outside the chosen closed core and also
ignoring higher lying unoccupied levels not relevant for the
properties of the nuclei of interest (see Fig.
\ref{fig:interact-SM}).

The single-particle energies (SPE) for the calculations may be taken
from a modern self-consistent mean-field calculation or from
experimental single-particle energies. The residual interaction
between the valence particles is taken into account by means of
two-body matrix elements (TBME) $\langle j_1,j_2,J | V
|j_3,j_4,J\rangle$ for all possible combinations of oribtals $j_i$
in the model space\footnote{In principle also three-body
interactions have to be taken into account and their influence is a
topic of current investigations \cite{Pie93,Ots09a}.}, where $J$ is
the total angular momentum to which the two single-particles angular
momenta couple. An effective interaction $V$ for a specific model
space has to include all possible combinations of orbitals $j_i$ and
generally depends on spin and isospin. From the SPEs and TBMEs the
excitation spectra of all nuclei for a given configuration space can
be calculated. For example, in the sd shell, involving the
$1s_{1/2}$, $1d_{3/2}$, and $1d_{5/2}$ levels and thus neutron numbers
N=8-20 and proton numbers Z=8-20, the SPE of 3 levels  and 63 TBME
have to be known \cite{Wil84}. These are then used as input for the determination
of the energies and wave-function of about $10^6$ nuclear states in
the mass region $A=17-39$.

TBMEs can in principle be derived from the bare nucleon-nucleon
interaction, which has been measured in nucleon-nucleon scattering,
using the so-called G-matrix theory \cite{Bru55,Gol57}. However, in
order to account for changes of the nucleon-nucleon interaction in
the nuclear medium some adjustments, e.g. incorporating short-range
repulsion and core polarization, have to be included. In addition
the extracted effective interaction has typically to be modified
empirically to fit to existing data (see Refs.
\cite{Cor09,Hjo95,Bro01,Dea04,Cau05} for extensive reviews).

\subsection{Monopole shift of the single-particle energies}\label{sec:monopole}

A central interaction $V(|\vec{r}_1-\vec{r}_2|)$ that depends only on the distance $|\vec{r}_1-\vec{r}_2|$
between the two particles can be expanded within a complete set of functions, e.g. Legendre
polynomials $P_k(\cos{\theta_{12}})$:
\begin{equation}
V(|\vec{r}_1-\vec{r}_2|) = \sum_k \nu_k(r_1,r_2)P_k(\cos{\theta_{12}}),
\end{equation}
with  coefficients $\nu_k(r_1,r_2)$ and $\theta_{12} = \sphericalangle (\vec{r}_1,\vec{r}_2)$.
The most important multipole orders relevant for the shell structure
in nuclei are the monopole component and the quadrupole component.
The latter plays an important role in driving the system to
quadrupole deformed shapes, which will be discussed later on.

The average of the interaction over all
directions
\begin{equation}
V_{j_i,j'_k} = \frac{\sum_J (2J+1) \langle j_i,j'_k,J |V|j_i,j'_k,J \rangle}{\sum_J (2J+1)}
\end{equation}
is equal to the monopole component of the interaction between two nucleons ($i,k$ denote if they are
proton $\pi$ or neutron $\nu$) in orbitals with angular momenta $j$
and $j'$, respectively, and is of particular importance for the discussion of
changing shell structure since it has a direct influence on the
energy spacing of the single-particle levels.

The effect of the monopole component can be seen best when looking
at the so-called effective single-particle energies (ESPE)
\cite{Pov81,Uts99}. For a specific level the ESPE is defined as the
separation energy of this orbit calculated with the bare SPE of the
configuration space and including the effects of the monopole
interaction with all other levels in the configuration space. Thus
the spacing between the ESPEs defines the energies for the
excitation of individual nucleons, and thus the effective shell
gaps. It should be pointed out here that the monopole interaction is
not the only relevant aspect of the residual interaction. However,
its influence on the ESPE as a function of proton or neutron number
is one of the most decisive contributions with respect to changing shell
gaps.

\begin{figure}
\begin{center}
\includegraphics[width=\textwidth]{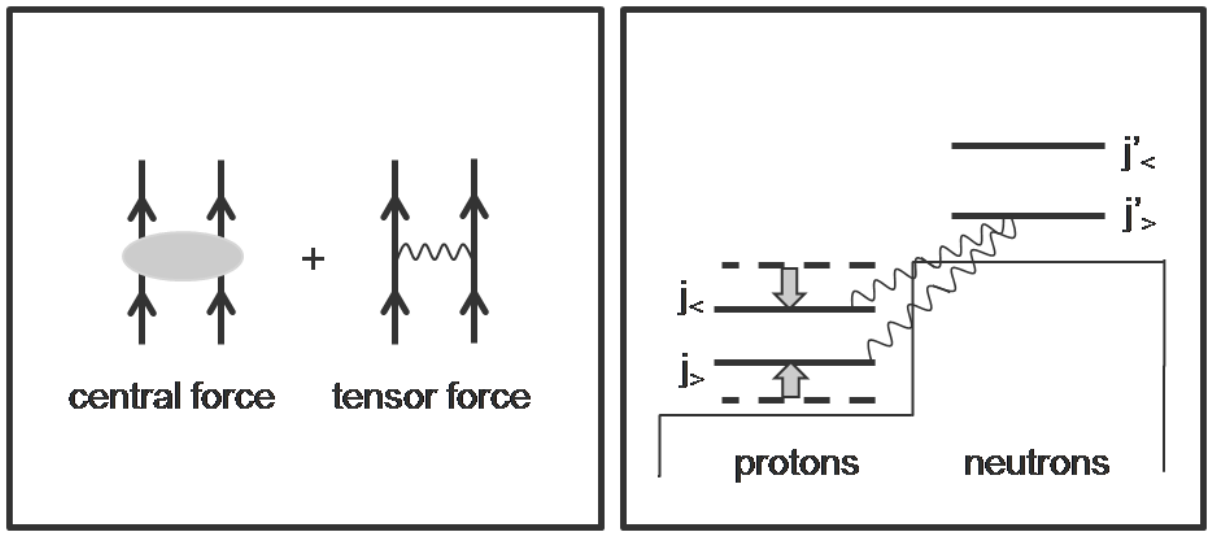}
\caption{{\it Left:} Basic components of the monopole interaction, a Gaussian central force and a tensor force.
(adopted from \protect\cite{Ots10}). {\it Right:} Schematic effect of the monopole interaction
between a $j'_>$ neutron orbital and $j_<$ and $j_>$ proton orbits.}%
\label{fig:monopole}
\end{center}
\end{figure}

It was recently pointed out by Otsuka and collaborators
\cite{Ots01,Ots05,Ots06,Ots10} that various changes in the shell
structure are caused by the monopole effect of the tensor force, an
important component in the free nucleon-nucleon interaction  which
had not been considered explicitly before in shell-model
calculations. The tensor force is one component of the residual
interaction and results from the $\rho$ and $\pi$ meson exchange
term of the nucleon-nucleon interaction. The important realization
was that in particular the monopole effect of the tensor force
between protons and neutrons depends on whether the orbits of these
valence nucleons are $j_>=\ell +s$ or $j_< = \ell -s$. The effect of
this interaction component is that if protons and neutrons are in
$j_<$ and $j_>$ or vice versa the single particle energies are
lowered (more bound), while in case that both are in $j_<$ or both
are in $j_>$ orbits the single particle energies are raised (less
bound), as indicated in Fig. \ref{fig:monopole}. Otsuka and
collaborators also showed that the evolution of shell structure in
exotic nuclei could be described by a so called monopole-based
universal interaction $V_{MU}$ \cite{Ots10}, consisting of a
Gaussian central force, containing many complicated processes
including multiple meson exchanges, and a tensor force comprised of
$\rho$ and $\pi$ meson exchange, as schematically indicated in the
left of Fig. \ref{fig:monopole}.

% Figure with j> and j< effects.

\begin{figure}
\begin{center}
\includegraphics[width=85mm]{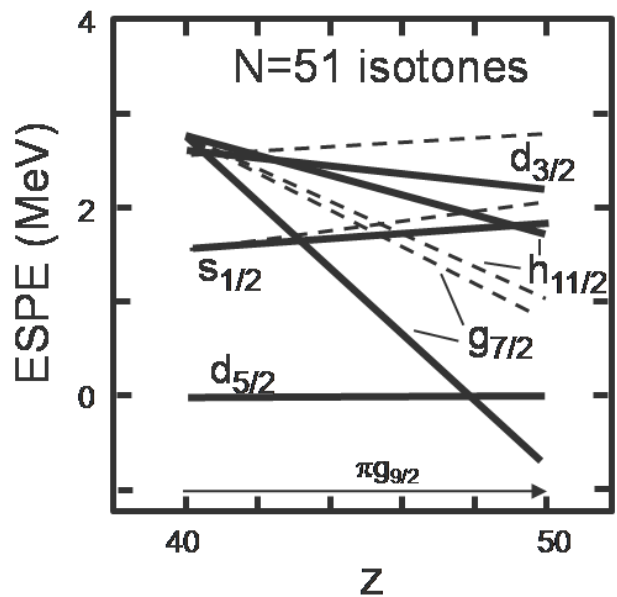}
\caption{Evolution of the neutron effective single particle energies (ESPE)
for the N=51 isotones as the proton $\pi g_{9/2}$ level is filled
(adopted from \protect\cite{Ots10}). The dashed lines are for the central force only,
while the solid lines include both the central force and the tensor force.}%
\label{fig:N51isotones}
\end{center}
\end{figure}

The effect of the tensor force can be seen in Fig.
\ref{fig:N51isotones} where the effective single particle energies
for neutrons are shown for the N=51 isotones as a function of proton
number Z. The figure shows how the EPSEs for the neutron $\nu 3s_{1/2}$,
$\nu 2d_{3/2}$, $\nu 1g_{7/2}$, and $\nu 1h_{11/2}$ levels change relative to
the $\nu 2d_{5/2}$ level as protons are added to the $\nu 1g_{9/2}$ orbital.
The dashed lines show the effect of the monopole part of a Gaussian
central force only, while the solid lines include the effect of the
monopole part of the tensor force. One can see the dramatic effect
on the $j_<$ neutron $\nu 1g_{7/2}$  orbital as its $j_>$ spin partner
proton $\nu 1g_{9/2}$ orbital is filled between Z=40 and 50. At the same
time the energy of the $j_>$ neutron $\nu 1h_{11/2}$ orbital is raised
by the repulsive effect of the tensor force. The effect of the
tensor force is particularly strong between spin-partner orbits due
to the fact that the radial wave-functions are the same and thus the
short range interaction can have maximum effect.

I will discuss the role of this interaction in the evolution of
shell structure and the occurrence of new magic numbers in section
\ref{sec:changing-shells}.

\section{Experimental probes of magic numbers}

In this section I want to briefly review some of the experimental
methods that enable us to explore the evolution of shell structure
and magic numbers in exotic nuclei. One may ask, why not simply measure
the single-particle energies. However, as will be discussed below,
such experiments are quite challenging and it is often easier to
resort to other data, which are more readily available, such as
measuring nuclear masses or investigating the excitation energies
and collectivity in even-even nuclei. However, first of all it is
necessary to briefly explain how exotic nuclei are produced and made
available for experiments.

\subsection{Production of exotic nuclear beams}
In the production of exotic nuclear beams one generally
distinguishes between two methods.

The first method, called Isotope Separation On Line (ISOL) method
employs light ion beams that induce spallation, fission, or
fragmentation reactions in a primary target, from which the
short-lived reaction products only escape by diffusion. They are
then singly ionized using e.g. surface ionization or element
specific laser ionization techniques. The ions are extracted from
the primary target and can be delivered to low-energy experiments,
such as a beta-decay station, a laser-spectroscopy set-up, or a
Penning Trap for mass-measurements. Alternatively, they can be
injected into a secondary post-accelerator for acceleration to
energies of around 2-10 MeV per nucleon, which are energies on the
scale of the Coulomb barrier between the nuclei. These
Coulomb-energy beams are used for inelastic scattering experiments
or nucleon transfer reactions, which are discussed below. Since the
diffusion out of the production target depends on the chemical
properties, ISOL facilities cannot provide secondary beams of all
elements. Also, the diffusion takes some time, limiting the reach of
the ISOL methods to exotic nuclei with half-lives of at least some
10 ms. The most important ISOL facilities worldwide are REX-ISOLDE
at CERN, HRIBF at ORNL and ATLAS at ANL in the U.S.A., ISAC at
TRIUMF in Canada, and SPIRAL at GANIL in France.

The second method relies on the  production of exotic nuclei by the
fragmentation or fission of a high-energy (50-1000 MeV per nucleon)
heavy ion beam in a production target, from which the reaction
products emerge with beam velocity. The nuclei of interest are
separated and identified by means of a large acceptance magnetic
fragment separator. The separated nuclei are used for secondary
reaction experiments at intermediate and relativistic energies
(40-1000 MeV per nucleon)  or can be stopped for decay studies. They
can also be injected into a storage ring for mass-measurements.
Since the separation technique does not depend on chemical
properties all elements are accessible in in-flight experiments.
Also nuclei with half-lives as low as a few hundred ns can be used
for experiments since the nuclei move at a large percentage
(40-80$\%$) of the speed of light. Current world-leading in-flight
facilities are the NSCL at MSU in the U.S.A., GSI Darmstadt in
Germany, GANIL in France, and RIBF at RIKEN in Japan. With the FAIR
Project at GSI Darmstadt with a planned begin of operations around
2017/18 the reach of the current facility towards more exotic nuclei
will be significantly extended.

NSCL and FRIB, with a planned startup around 2017, at MSU in the U.S.A. plan to combine the in-flight
production and separation with a gas-stopping cell from which ions
can be extracted and fed into a post-accelerator for secondary
experiments at Coulomb-barrier energies.

\subsection{Masses, binding energies, separation energies}
One of the first indications for the occurrence of a shell gap can
come from the mass of the nucleus, which can today be measured even
for nuclei with production rates of a few per second and half-lives
in the millisecond range.

The most precise technique to measure masses employs Penning Traps
and can achieve an accuracy of 10 keV and better \cite{Bla06,Bla10}.
As do all high precision measurements today, these experiments rely
on a frequency measurement. The ions with charge $q$ and mass $m$
are stored in a Penning trap by the superposition of a strong
homogeneous magnetic field $B$ for radial confinement and a weak
static electric field for axial trapping. The electric field is
applied to hyperbolical electrodes with axial symmetry around the
B-field direction. The electrode is fourfold segmented and the ions
are excited by a radio-frequency applied to the pairs of segments.
In case the applied radio-frequency is equal to the cyclotron
frequency $\omega_c = B\cdot q/m$, maximum energy is transferred to
the ion. When the ion is extracted from the trap its time-of-flight
to a detector is measured. Measuring the time-of-flight as a
function of the applied radio-frequency reveals a resonance pattern
with a minimum time-of-flight for the cyclotron frequency of the
ion, from which the mass can be extracted.

In an alternative approach masses of nuclei can be measured in a
storage ring \cite{Fra87,Fra08} in which they circulate with about
$10^6$ revolutions per second, with the revolution frequency
depending on $q/m$ of the ion. In order to suppress effects on the
revolution frequency resulting from the velocity spread of the ions
one can perform the experiment in two different ways. One technique
relies on the operation of the storage ring in a so called
isochronous mode, where the circulation time of the ions does not
depend on their velocity \cite{Hau00}. A measurement of the
time-of-flight using special detectors allows one to measure the
revolution frequency, which now only depends on $q/m$. The other
method relies on cooling of the ions in an electron cooler to reduce
their velocity spread to a negligible value. In this case the
revolution frequency can be measured by the Schottky noise signal
induced by the ions in an electrode installed in the ring, which is
then analyzed by applying a Fast Fourier Transformation
\cite{Rad00}.

The measured masses can be used to investigate the shell evolution
due to the fact that nuclei with a closed shell configuration are
bound more strongly compared to nuclei with one or a few extra
nucleons. Thus one can see indications for a shell closure either by
an enhanced binding energy of the nuclei with closed shells or by a
significantly reduced separation energy for additional nucleons. For
example, the neutron separation energy $S_n$ for a nucleus with $Z$
protons and $A$ nucleons is given by
$$
S_n  =  BE(Z,A) - BE(Z,A-1)\\
     =  [M(Z,A-1)-M(Z,A)+M_n]\cdot  c^2.
$$
Here $BE(Z,A)$ is the binding energy of a nucleus with mass M(Z,A)
and $M_n$ is the neutron mass.

\begin{figure}[ht!]
\begin{center}
\includegraphics[height=150mm]{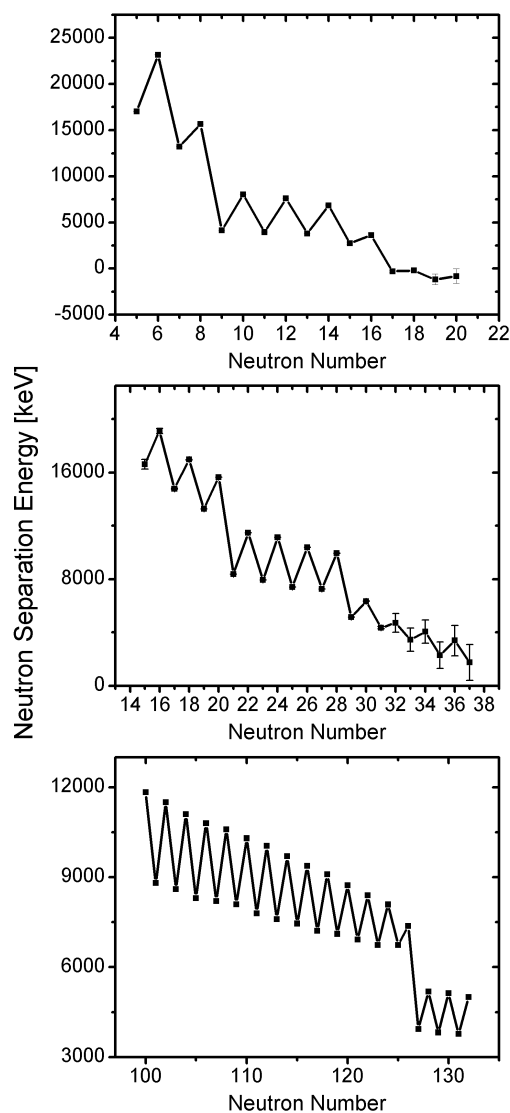}
\caption{Neutron separation energies $S_n$ for oxygen (top), calcium (middle),
and lead (bottom) isotopes taken from \cite{AME03}, respectively.}%
\label{fig:separation}
\end{center}
\end{figure}

Figure \ref{fig:separation} shows neutron separation energies $S_n$
for oxygen, calcium, and lead isotopes as a function of neutron
number. One can clearly see the jumps in the separation energies for
the well known classic magic numbers N=126 in lead, N=20, 28 in
calcium and N=8 in oxygen.

%(FIG Separation energy indicating shell effects - classic as well as Oxygen)

However, local shifts in the binding energies or separation energies
can sometimes also be the result of other causes, e.g. the
competition and mixing of different nuclear shapes at low energies. Therefore,
the ground state properties alone are not always sufficient to
establish a new shell gap or the disappearance of a known shell gap.
However, important additional information can be gained from the study of
excited states.

\subsection{Excited states and collectivity}
Additional evidence for a shell closure can come from low lying
excited states in nuclei. For example, in nuclei with one particle
or hole outside a closed shell the excited states directly provide
information on the single-particle states.

For exotic nuclei with production rates between about 1 per second
and 1 per hour it is possible to study the gamma-decay of excited
states populated either in the radioactive decay of the mother
nucleus or in the decay of a long-lived isomeric state populated in
the production reaction of the exotic nucleus. In either case the
subsequent gamma-decay can be observed establishing the sequence of
excited states and their electromagnetic decay. However,  the total
angular momentum and parity of the emitting state can be established
only by measuring angular distributions or correlations of the
gamma-radiation depopulating this level, which requires significant
statistics and thus is often not possible in exotic nuclei. Lacking
this information, the observation of states in nuclei with only one
particle or hole outside a closed shell is only partially helpful
since not all quantum numbers of the relevant states can be
established.

However, even-even nuclei, with an even number of protons and
neutrons, enable a more simple access to the investigation of shell
structure.  The well known pairing interaction, that couples the
angular momenta of like nucleons pairwise to total angular momentum
0, leads to the lowering of the energy of $0^+$ states in even-even
nuclei. As a result all even-even nuclei  have a spin and parity
$J^{\pi}=0^+$ ground state and a simple  low-energy excitation
spectrum with a $2^+$ first excited state in almost all cases.
Therefore, an explicit measurement of angular momentum and parity of
this first excited state in even-even nuclei is basically not
needed. At the same time its energy is a good indicator for the
existence of a shell closure, as can be understood by the following
considerations.

In order to generate an excited $2^+$ state in a doubly magic
nucleus at least one nucleon has to be excited across the shell gap,
since all single-particle levels are fully occupied and coupled to a
total spin $J=0$, such that a simple recoupling of angular momenta
to $J=2$ is not possible. The cross-shell excitation costs a large
amount of energy, leading to a high excitation energy of several MeV
for the $2^+$ state. In even-even nuclei with one closed and one
open shell it does cost significantly less energy to produce this
excited state, namely the energy to break a pair of nucleons in the
open shell, about 1-1.2 MeV.

\begin{figure}
\begin{center}
\includegraphics[width=85mm]{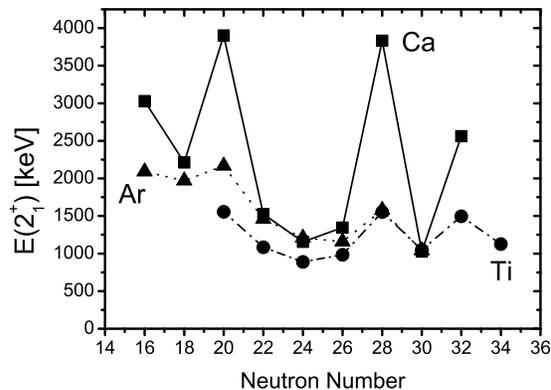}
\caption{Experimentally known energies of the first excited $2^+$ states in the even-even Ar, Ca, and Ti isotopes}%
\label{fig:2plus}
\end{center}
\end{figure}

Fig. \ref{fig:2plus} shows the energies of the lowest $2^+$ states
in the even-even Ar, Ca, and Ti isotopes. The high excitation
energies at the classic neutron magic numbers 20 and 28 are clearly
visible for all three elements, while the highest energies occur in
case of the doubly magic nuclei $^{40}$Ca and $^{48}$Ca. The high
excitation energy of the $2^+$ state in $^{52}$Ca and $^{54}$Ti is
due to the filling of the $\nu 2p_{3/2}$ orbital, which consists of
a sub-shell.

If both shells for protons and neutrons are open it is energetically
less costly to produce excited $2^+$ states since the long range
quadrupole-quadrupole interaction between the valence protons and
neutrons allows for surface vibrations (near the closed shells)
($\approx$ 0.6 MeV) and a permanent quadrupole deformation (mid
shell) with low lying rotational excitations ($\approx$ 0.05-0.4
MeV). Rotational and vibrational excitations are so-called
collective excitations where several/many nucleons are coherently
contributing to the overall motion of the nucleus, in one case
surface vibrations of quadrupole shape \footnote{Also octupole shape
vibrations are possible but will not be discussed further here.}
and in the other case the rotation of a
quadrupole shaped charge distribution. In both cases a large
(collective) transition matrix element $\langle 2^+ | \hat{Q} |
0^+\rangle$ exists for the electromagnetic quadrupole operator
$\hat{Q}=r^2Y_{20}$. The value of this matrix element is orders of
magnitude larger than for the case where a single-nucleon is moved
by an electric quadrupole interaction from one single-particle level
to another one.

Therefore, the strength of the electromagnetic transition between
the $0^+$ ground state and the first excited $2^+$ state in
even-even nuclei provides another indication for shell closures. The
strength is often measured in terms of the reduced transition
strength $B(E2;0^+ \rightarrow 2^+) \propto |\langle 2^+ | Q |
0^+\rangle |^2$. Large B(E2) values are found for collective nuclei
near mid-shell while very small B(E2) values will be found at shell
closures.

Aside from gamma-ray spectroscopy in decay experiments one can gain
access to the first excited $2^+$ state in even-even nuclei via
inelastic scattering either using electromagnetic (Coulomb)
excitation or using the strong-interaction. In the Coulomb
excitation of an exotic nuclear beam the projectile is scattered on
a stable high-Z nucleus in order to ensure a strong Coulomb
interaction or, in other words, the availability of many virtual
photons for the excitation. Coulomb excitation is a
very well established technique \cite{Ald56,Ald66} and theoretically
under control since the interaction and the reaction kinematics are
well known. It has been adopted for the study of exotic nuclei at
beam energies below the Coulomb barrier (see e.g.
\cite{Bro91,Osh92,Nie05}) as well as at energies well above the
Coulomb barrier between projectile and target nucleus, often coined
intermediate energy or relativistic Coulomb excitation
\cite{Mot95,Gla98,Gad08}. In either case the scattered projectiles
are observed in coincidence with the characteristic gamma-ray
transition from the deexcitation of the first excited $2^+$ level.

The advantage of the Coulomb excitation method is that, in addition
to the observation of the energy of the $2^+$ state, the excitation
cross-section can be used to obtain information on the
electromagnetic transition matrix element connecting ground state
and the $2^+$ state. However, one needs to ensure that the
interaction is of electromagnetic nature only, which is the case if
the distance between the scattering partners stays well above the
range of the strong interaction. At beam energies significantly
below the Coulomb barrier between the scattering partners this is
easily achieved for all scattering angles, and thus the technique is
sometimes called ''safe'' Coulomb excitation. Such experiment are
performed at various ISOL facilities. For the "safe" Coulomb
excitation at energies below the Coulomb barrier one is restricted
to thin targets ($\approx$ mg/cm$^2$) and thus experiments need beam
intensities of $\approx 10^4$ particles per second or above.

For the Coulomb excitation with high beam energies at in-flight
facilities the selection of excitation via pure Coulomb interaction
is basically achieved by selecting very small scattering angles,
which ensures that the reaction products did not approach each other
too closely. The relativistic Coulomb excitation has the advantage
that thick targets ($\approx$ g/cm$^2$) can be used. This allows one to
perform experiments with secondary beam intensities of only
$\approx$ 10 particles per second.

Aside from the pure electromagnetic excitation one can alternatively
use the inelastic scattering of the exotic projectile on a target
containing Hydrogen (protons) or Helium (alpha particles). Since in
this case the projectile is dominantly excited by the strong
interaction, one can obtain different information on the excited
states. The electromagnetic excitation is only sensitive to the
proton contribution while inelastic proton scattering, for example,
is also sensitive to the neutron contribution of the transition. See
Ref. \cite{Gad08} and references therein for further details.

\subsection{Energies and occupations of single-particle levels}

While masses, $2^+$ energies, and transition matrix elements provide
important insights as to the existence of a shell closure, it is
important, if possible, to obtain direct information on the energies,
quantum numbers, and occupations of the single-particle levels above
and below a shell closure. In order to obtain this information one
needs to measure the quantum numbers and energies of the
single-particle energies in nuclei with or near closed shells. This
can be done by so-called transfer reactions, in which one or few
nucleons are transferred between a projectile and target nucleus in
a peripheral collision where both nuclei just touch. Such
experiments have been performed for
more than 40 years at stable beam accelerators using
light projectiles, like protons, deuterons, $^3$He, or $^4$He.
For example, a neutron can be placed in the
empty single-particle levels outside a closed shell nucleus, e.g.
$^{48}$Ca, using a deuteron beam and detecting the residual proton
from this $(d,p)$ reaction. In this case the neutron can be placed in the
$\nu 2p_{3/2}$, $\nu 1f_{5/2}$, or $\nu 2p_{1/2}$ levels in $^{49}$Ca. The
energies and angular momentum quantum numbers of the populated
levels can be measured by detecting energies and angular
distributions of the residual protons, e.g. with a high resolution
magnetic spectrograph. Such experiments can be performed within a
day or so at beam intensities of $10^{10-12}$ particles per second
(pps) and with beam energies of around 10 MeV per nucleon or higher.

For short lived exotic nuclei this technique has to be modified by
applying inverse kinematics. At various ISOL facilities,
post-accelerated exotic nuclear beams are used to induce (d,p)
neutron transfer reactions in deuterated polyethylene foils to study
the neutron single-particle energies far away from stability. In
such experiments the protons from the target are detected in silicon
detector arrays surrounding the target, measuring energies and
emission angles. Due to the energy loss of the heavy-ion beam in the
target and due to kinematic effects resulting from the reversed
scattering geometry the energy resolution in such experiments is
rather poor. However, one can use gamma-ray detection in coincidence
with the detected protons to select excited states of interest.
Experiments of this kind need beam intensities of at least $10^4$
particles per second.

Another way to probe the energies, quantum numbers, and occupations
of single-particle levels in a doubly magic nucleus is the so-called
knockout reaction \cite{Han03,Gad08} with relativistic beams at
in-flight facilities. In these experiments a fast secondary exotic
nuclear beam is directed onto a thick ($\approx$ g/cm$^2$) carbon or
beryllium target at typical energies of 50-1000 MeV per nucleon.
Due to the use of thick targets, which is enabled by the high beam
energies, experiments can be performed with secondary beam rates of
as low as 1 particle per second. In this case each incoming particle
can be identified in its mass and charge through
measurements of the magnetic rigidity, time-of-flight (providing
velocity), and energy loss, which depends on the nuclear charge $Z$.
In the knockout reaction a nucleon at the surface of the projectile
is knocked out of its orbital by a peripheral collision with the
target nucleus. The residual nucleus with one less nucleon is again
guided through a magnetic spectrometer that on one hand allows for
an event-by-event identification of the knockout reaction products\footnote{Aside
from the knockout reaction various other
reactions occur.} and on the other hand yields a precise measurement
of the momentum change of the projectile like residue compared to
that of non reacted projectiles.

The change in momentum of the residue is, due to momentum
conservation,  a direct measure of the momentum of the
single-particle that was removed from its orbital in each individual
reaction. For many knockout reactions on the same single-particle
orbital one obtains a momentum distribution of this particular
orbital. By means of a Fourier transformation this distribution is
related to the spatial distribution of the radial wave-function of
the removed nucleon, which varies for different orbital angular
momenta $L$ of the involved single-particle level. Thus the measured
momentum distribution for an individual level is characteristic for
the orbital angular momentum of the single-particle level from which
the nucleon is removed. The cross-section for the knock-out reaction
can be reliably calculated using Glauber theory \cite{Han03} and can
be used as a measure of the occupation of the orbital, which can be
compared with the predication of shell model calculations.

Thus, direct reactions like transfer and knockout reactions are
essential tools to obtain detailed information on the energies of
the single-particle levels and on the number of particles occupying
these levels.

\section{Changes of shell structure far away from
stability}\label{sec:changing-shells}

After having provided a rough overview on the theoretical basics of
the shell model and the most important experimental observables used to
probe the occurrence of shell gaps it is now time to turn towards
the recent findings of changing shell structure in exotic nuclei.
The discovery of such changes were only possible due to the recent
developments in providing energetic beams of exotic nuclei and
enabling the investigation of nuclei further and further away from
stability. Through these studies it has been possible to
significantly extend our knowledge of the evolution of nuclear
structure with isospin. Hereafter, I only can highlight a few of the
recent developments explicitly, while more information can be found
in various reviews, e.g. Refs. \cite{Aum05,Gad08,Sor08}.

On the theoretical side a number of studies over the years have
suggested that shell structure may change far away from stability
for a number of reasons \cite{Ton78,Hae89,Dob94,Pea96,Lal98}. Among
the various mechanisms for shell changes in the mean-field potential
the reduction of the spin-orbit interaction with increasing neutron
excess has been discussed extensively. From a phenomenological point
of view the spin-orbit interaction $V_{LS}\propto dV(r)/dr$ may be
expected to decrease for a more diffuse nuclear potential $V(r)$
that may result from the weaker binding of the additional neutrons.

Alternatively, the monopole shifts originating from the residual
interaction of the valence nucleons, as discussed in section
\ref{sec:monopole}, may cause the shell structure to change
dependent on the occupied single-particle levels.

In this section I want to highlight a few examples of regions in the
nuclear chart where major changes in the shell structure have been
observed or are currently debated. The evolution of all major shell
closures has been extensively discussed in a recent review
\cite{Sor08}, to which the reader is referred to.

I will start in the region of light elements between oxygen and
calcium, where the shell structure is modified by the residual
interaction. Afterwards I will discuss the situation in heavier nuclei, e.g. near
$^{132}$Sn, where possible evidence for modifications of the shell
structure has been discussed and changes of the spin-orbit
interaction as well as effects of the residual interaction have been
put forward in this context.

\subsection{Light nuclei}

\begin{figure}
\begin{center}
\includegraphics[width=\textwidth]{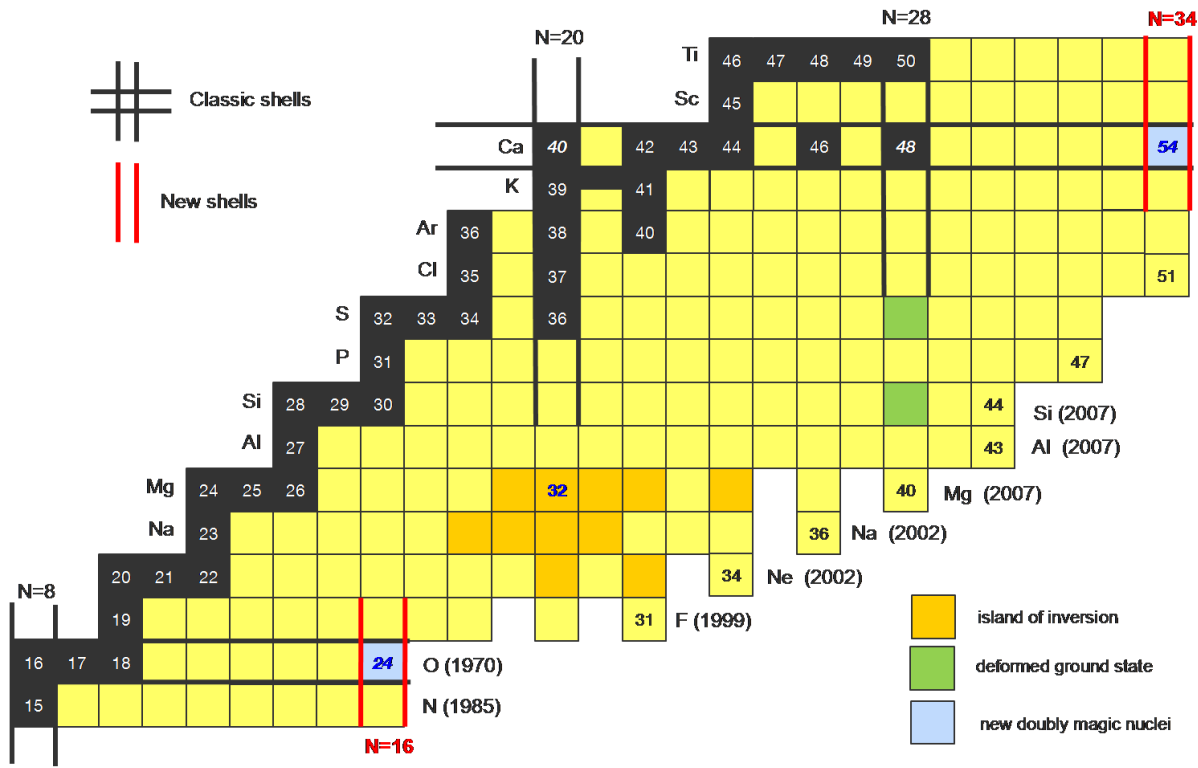}
\caption{Nuclear chart between nitrogen (Z=7) and titanium (Z=22) indicating classic and new shell closures
as well as the breakdown of classic shell closures.}%
\label{fig:new-shells}
\end{center}
\end{figure}

Fig. \ref{fig:new-shells} shows the neutron-rich side of the nuclear
chart between Z=7 (nitrogen) and Z=22 (titanium) indicating the
stable isotopes in black. The unstable nuclei are mostly indicated
in yellow, aside for those nuclei where changes in shell structure
are indicated by different colors:
\begin{itemize}
\item light blue for the new doubly magic nuclei $^{24}$O (N=16) and $^{54}$Ca
(N=34) (see below).
\item orange for the Island of Inversion (see below), where the N=20 shell
closure has eroded and deformed configurations based on the
pf-shell above N=20 dominate the ground state configurations of
nuclei near N=20.
\item green for the N=28 isotones $^{42}$Si and $^{44}$S where experimental
data strongly indicates that their ground states are deformed
and the N=28 shell gap disappears (see e.g. \cite{Sor08,Gad08}
and references therein).
\end{itemize}

%Fig of nuclear chart and dripline a la Otsuka with dates of discovery of last isotopes...

Also listed are the years when the most-neutron-rich isotope of each
element was discovered. This indicates our current knowledge of the
so called neutron drip-line. The evolution with time, in
particular in the last 20 years or so is closely related to the
aforementioned developments of radioactive beam facilities.
The development of theoretical and experimental predictions went hand
in hand. While in some cases, e.g. in case of the Island of Inversion
the experimental evidence for the breakdown of the N=20 gap came first,
in other cases, e.g. the prediction of the N=34 shell closure, theoretical
predictions motivated experimental activities.

One can also see that the neutron drip line follows an almost linear
dependence, aside from odd-even effects. The oxygen isotopes represent a
major deviation from this smooth trend with $^{24}$O (N=16) being
the last bound oxygen isotope, despite the fact that the simple
shell model expectation would tell us that $^{28}$O with  20
neutrons in a closed shell should be favored in terms of nuclear
binding. On can also see that if one adds only one additional proton
suddenly six more neutrons can be bound in the fluorine isotopes
where the last bound isotope is $^{31}$F.

\subsubsection{Experimental evidence for the $N=16$ magic number}

The first evidence for the $N=16$ magic number in oxygen came from
an evaluation of neutron separation energies on the basis of
measured masses. From the neutron separation energies of the oxygen
isotopes in Fig. \ref{fig:separation} it is apparent that the
nucleus $^{28}$O with N=20 is not bound anymore, despite its classic
magic neutron number, which should provide additional stability. To
the contrary, one can notice a jump in the oxygen separation
energies at N=16. This surprising observation was taken as the first
indication that instead of N=20 there may be a new shell closure at
N=16 for oxygen \cite{Oza00}.

In a recent experiment at the National Superconducting Cyclotron
Laboratory (NSCL) at Michigan State University, USA, the energy of
the first exited $2^+$ state in $^{24}$O could be deduced
\cite{Hof09}. Since this state, produced by the removal of a proton
and a neutron from an exotic beam of $^{26}$F, is unbound, it decays
promptly to the ground state of $^{23}$O under emission of a
neutron. In this experiment $^{26}$F nuclei were produced in a first
step by fragmenting a $^{48}$Ca beam on a Be target and separating
and identifying the $^{26}$F fragments in the A1900 fragment
separator. The $^{26}$F fragments were then sent onto a secondary Be
target so that $^{24}$O was produced by the removal of a proton and
a neutron from the $^{26}$F beam. By identifying $^{23}$O and the
decay neutrons at the same time it was possible to reconstruct the
energy spectrum of excited states in $^{24}$O. The high energy of
4.72 (11) MeV of the $2^+$ state provides clear evidence of the
spherical nature of $^{24}$O, as expected for a doubly magic
nucleus.

Further evidence for the doubly magic nature of $^{24}$O comes from
its single-particle structure, which was studied in a knockout
experiment \cite{Kan09} at the Fragmentseparator FRS \cite{Gei92} of
the GSI Helmholtzzentrum f\"ur Schwerionenforschung in Darmstadt,
Germany. In this experiment it was found that the measured momentum
distribution for a neutron removal from the ground state in $^{24}$O
showed a pure $L=0$ character. This proved that the last two
neutrons in $^{24}$O occupied only the $\nu 2s_{1/2}$ single-particle
level as predicted by a shell model calculation including the new
shell closure at $N=16$. If there were no new magic number $N=16$ in
the oxygen isotopes the $\nu 1d_{3/2}$ level would have been close to
the $\nu 2s_{1/2}$ level and a substantial contribution of $L=2$
would have been measured in the momentum distribution, which was not
the case.

All experimental evidence available on $^{24}$O and its neighboring
isotopes clearly support the fact that N=16 is a new magic number
(see also \cite{Sor08,Gad08,Jan09}).

\subsubsection{Effective Single Particle Energies for N=20}

On the basis of the theoretical knowledge concerning the effect of
the residual interaction on the single-particle energies the
disappearance of the N=20 and appearance of the N=16 magic number in
the oxygen isotopes can be explained.

\begin{figure}
\begin{center}
\includegraphics[width=85mm]{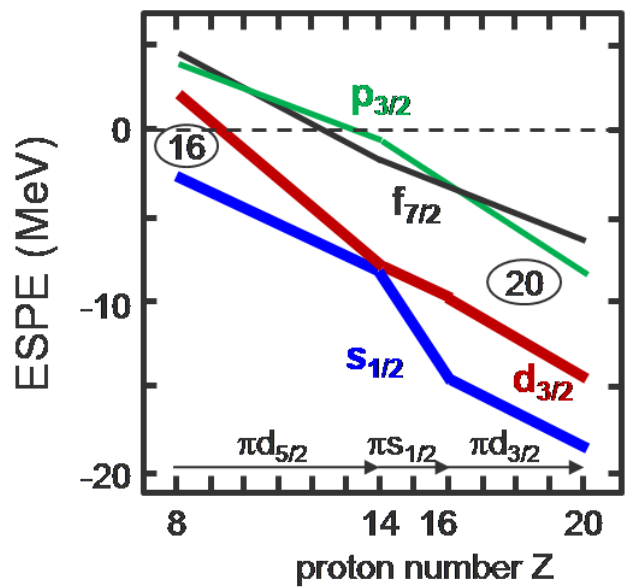}
\caption{Effective single particle energies (ESPE) for the N=20 isotones
between oxygen and calcium (adopted from \protect\cite{Ots10}).}%
\label{fig:N20-ESPE}
\end{center}
\end{figure}

 Figure \ref{fig:N20-ESPE} shows the effective single-particle
energies for the N=20 isotones, adopted from \cite{Ots10}, as a
result of the monopole component of the residual interaction
$V_{MU}$ with Gaussian central and tensor part (see Fig.
\ref{fig:monopole}). One can see the large N=20 shell gap for
calcium (Z=20), which decreases in size once protons are removed
from the $\pi 1d_{5/2}$ orbital below Z=14. At Z=8 a sizable N=16
shell gap is present. The starting point for this plot are the Z=8
SPE from the SDPF-M interaction \cite{Uts99}. Note that $^{28}$O is
unbound.

By adding protons in the $\pi 1d_{5/2}$ orbital the $\nu 1d_{3/2}$
orbital gets more bound and the classic N=20 shell gap is opening
up. The N=16 shell gap for Z=8 is thus due to the absence of the
attractive monopole interaction between the $\pi 1d_{5/2}$ the $\nu
1d_{3/2}$ levels.

The effect of the monopole interaction can be also seen when adding
a single proton into the $\pi 1d_{5/2}$ orbital. In this case there
is some additional binding of the $\nu 1d_{3/2}$ level due to its
interaction with the $\pi 1d_{3/2}$ level allowing for the binding of
six more neutrons all the way to $^{31}$F.

\subsubsection{The Island of Inversion}

The ESPE for the N=20 isotones in Fig. \ref{fig:N20-ESPE} show
another interesting feature for Z=10 and Z=12. For these proton
numbers there is no clear shell gap visible. The N=16 shell gap is
large only for oxygen and carbon (Z=6) and the N=20 shell gap is
well developed only for Z=14 and above. For Ne and Mg there is a
more even spacing of all the levels and in this case quadrupole
correlations can come into play since their strength is of the order
of the level spacing. In such a case quadrupole collectivity can
develop by a partial occupation of levels of the pf shell, which in
turn drive the ground states of the N=20 isotones $^{30}$Ne and
$^{32}$Mg into deformation. Due to the effective lowering of the
pf-shell orbits compared to the standard sd-shell levels, one talks
about an inversion of the single-particle levels. The nuclei around
$^{30}$Ne and $^{32}$Mg, which show ground state deformation and thus
a breakdown of the N=20 shell, are considered to be part of this
so-called {\it Island of Inversion} \cite{War90}. The first proof of
the role of deformation for these nuclei came from the Coulomb
excitation of an exotic $^{32}$Mg beam at the RIPS fragment
separator at RIKEN, Japan \cite{Mot95}. However, until today the
exact borders of the island of inversion are still not delineated
completely and there are still intensive experimental and
theoretical efforts underway to exactly understand the evolution of
shell structure and the competition of spherical and deformed
configurations in this mass region (see e.g.
\cite{Gad08,Sch09,Doo09,Kan10} and references therein).

\subsubsection{The new magic number N=34}
When Honma et al. \cite{Hon02} introduced the new effective shell
model interaction GXPF1 for the nuclei in the pf-shell, which
included the effect of the monopole component of the tensor
interaction, they predicted that the first excited $2^+$ energy in
the N=34 nucleus $^{54}$Ca would be as high as for the doubly magic
$^{40,48}$Ca nuclei. At the same time, shell-model calculations
using the well established KB3G interaction
\cite{War90,Pov01,Cau02,Pov05} support a $N=32$ shell closure, which
is experimentally well established, but not a $N=34$ shell closure.

The prediction of a new doubly magic shell closure for $^{54}$Ca
triggered a large number of experimental and theoretical
investigations. So far, $^{54}$Ca has been out of reach and thus the
experimental investigations have concentrated on the nuclei in its
vicinity. A number of studies have been performed on neutron rich
$Z=20-24$ nuclei using $\beta$-decay as well as multi-nucleon transfer,
Coulomb excitation, and knockout reactions of radioactive ion
beams (see \cite{Gad08,Mai09} and references
therein). The level schemes obtained for these nuclei compare
favorably with shell-model calculations using the GXPF1A interaction
and thus support the prediction of the new N=34 shell closure.
Clarity concerning this very interesting question will only be
obtained when  experiments can finally be performed on $^{54}$Ca
directly. This may be possible within the next few years at
the new Radioactive Ion Beam Factory (RIBF) at RIKEN.

\subsection{Shell quenching in heavier nuclei}

In the preceding chapters one of the main mechanisms for the change
of shell structure in neutron-rich nuclei has been discussed, namely
the shift of single-particle energies due to the monopole component
of the residual interaction. However, self-consistent mean-field
studies have also predicted changes in the shell structure, in
particular when approaching the
neutron-dripline \cite{Ton78,Hae89,Dob94,Pea96,Lal98}. The most
significant effect is the quenching of shell gaps near the dripline
due to the coupling of bound states to unbound continuum states
above the Fermi-energy, e.g. through pair scattering. In addition,
 an isospin dependence of the spin-orbit interaction has been
discussed. For increasing neutron-to-proton ratios this could lead
to a reduction of the splitting between
spin-orbit partner orbits, which is the main cause for the shell
gaps in medium mass and heavy nuclei. Thus a reduction of the
spin-orbit interaction, related to a more diffuse nuclear surface,
would also cause a shell quenching.

Experimental evidence for the quenching of the N=82 shell was
suggested by the observation \cite{Dil03} of a less than expected
binding in the N=82 nucleus $^{130}$Cd, just two neutrons below
doubly-magic $^{132}$Sn. The binding energy was determined by
measuring the beta-decay endpoint energy and thus the mass
difference to the daughter nucleus of known mass. The experiment was
performed at the ISOLDE radioactive beam facility at CERN, one of
the longest operating radioactive beam facilities with a large
number of elements available for experiments. However, recent
spectroscopic studies of the excited states in $^{130}$Cd at GSI
Darmstadt revealed that the excitation pattern can be well described
by SM calculations in which the N=82 shell gap has its full size
\cite{Jun07}. The N=82 nuclei below $^{130}$Cd are just coming into
the reach of RIB facilities and the next years will probably solve the puzzle
of possible shell quenching below $^{132}$Sn.

Another investigation in this mass region studied the evolution of
the energy spacing between the proton orbitals $\pi 1g_{7/2}$ and
$\pi 1h_{11/2}$ in the antimon (Sb) isotopes as a function of neutron
number. An increase of this spacing was observed \cite{Sch04} and is
interpreted as a lowering of the $\pi 1g_{7/2}$ and a rise of the
$\pi 1h_{11/2}$ orbital, which in turn means that the spacing between
the $\pi 1g_{7/2}$ ($\pi 1h_{11/2}$) orbital and its spin-orbit
partner orbital $\pi 1g_{9/2}$ ($\pi 1h_{9/2}$) is decreasing with
increasing neutron number. This observation was taken as a possible
indication of a reduction of the spin-orbit interaction with
increasing neutron number. However, Otsuka et al. \cite{Ots05,Ots06}
pointed out that the monopole part of the tensor interaction would
produce exactly the same effect, since within the studied range of
neutron numbers the $\nu 1g_{9/2}$ orbital is filled. The equivalent
effect for neutron orbitals in the N=51 isotones has been discussed
in section \ref{sec:monopole} and Fig. \ref{fig:N51isotones}.

In addition to the aforementioned studies, there is also an intense
effort to study the persistence of classic shell closures in exotic
nuclei. On the neutron-rich side the N=50 shell closure around
doubly-magic $^{78}$Ni \cite{Hos05,Wal07,Bar08,Hak08} and the N=126
below $^{208}$Pb \cite{Ste08,Ald09,Pod09}, which are both relevant
for the delineation of the r-process path, have become accessible to
first experimental studies. On the proton-rich side the doubly-magic
$^{48}$Ni has been produced for the first time \cite{Bla00}.
$^{48}$Ni lies near the proton drip-line and there is evidence that
it is unbound against two-proton emission from its ground state
\cite{Dos05} where the unbound protons are emitted via tunneling
through the Coulomb barrier. See Refs. \cite{Bla08,Woo97} for recent
reviews on decay studies at the proton dripline. Also the heaviest
N=Z doubly magic nucleus $^{100}$Sn and its immediate vicinity has
been reached experimentally and beta decay studies have recently
been performed with production rates of about 1 per hour
\cite{Baz08,Fae08,Hin10,Epp10}.

Another frontier of the investigation of shell closures in nuclei is
the question of which shell gaps determine the structure of the
superheavy nuclei and if the next spherical shell closure beyond Pb
(Z=82) is Z=114, 120, or 126. Here various theoretical models come
to different conclusions \cite{Sob07} and with tremendous
experimental efforts it has been possible to expand the nuclear
chart up to Z=118 using fusion reactions with intense stable ion
beams (see Ref. \cite{Hof00,Oga07} for recent reviews). While
apparently there is an increased stability of the heaviest, most
neutron-rich superheavy elements observed, the question of the
position and size of the next spherical shell closure remains open
so far.

\section{Conclusion}
In this review I have tried to summarize some of the main aspects of
nuclear shell structure and highlighted some of the main drivers of
shell modifications far away from stability.  Quite dramatic changes
of shell structure occur in the neutron-rich region of the nuclear
chart, in particular in light nuclei, as indicated in figure
\ref{fig:new-shells}. For example, the N=20 and 28 shell closures
break down and new magic numbers appear at N=16 for Z=6,8 and N=34
for Z=20. The discovery of such major changes to the magic numbers,
which were previously thought to be valid throughout the whole
nuclear chart, has lead to vigorous experimental and theoretical
activities to understand the origin of these changes and to make
predictions for those nuclei not accessible to experimental studies
yet. From the above discussion of nuclei near $^{132}$Sn one
realizes that the effect of the monopole interaction and any changes
of the spin-orbit interaction can contribute to the observed changes
of the single-particle structure at the same time and need to be
carefully disentangled. In addition, other effects can come into
play, such as changes of the nuclear pairing correlations,
collective correlations, or the coupling of bound states to unbound
continuum states.

The complexity of these issues is one reason why it is essential to
obtain multiple and detailed experimental observables on the
single-particle structure far away from stability. The development
of facilities for exotic nuclear beams have been an essential driver
for these studies. However, due to the difficulties in producing the
nuclei of interest and the low intensities of those exotic nuclear
beams, the experiments are very challenging and need ingenious
technological and methodological developments to succeed. It is this
combination of theoretical and experimental challenges, which make
the field of nuclear structure of exotic nuclei so interesting.

The radioactive beam facilities of the next generation like RIBF at
RIKEN (Japan), FAIR at GSI Darmstadt (Germany), and FRIB at MSU
(U.S.A.) as well as HIE-ISOLDE at CERN, SPIRAL2 at GANIL (France)
and possibly EURISOL will provide access to even more exotic nuclei,
in particular along the r-process path and all the way towards the
neutron dripline for medium-mass nuclei. With these facilities and
the immense theoretical efforts to develop a unified theoretical
framework for the description of the structure and dynamics of
nuclei the evolution of shell structure in atomic nuclei will be
possible to develop a coherent understanding of the nuclear many
body system and link it to the other facets of the phases and structures
of strongly interacting matter.

\subsection*{Acknowledgements}
I would like to thank J. Friese, R. Gernh\"auser, and D. M\"ucher
for the careful reading of this manuscript and many thought
provoking discussions.

\bibliographystyle{tCPH}
\bibliography{tCPHguide}

\begin{thebibliography}{10}
\bibitem{Bed02} P.F. Bedaque, U. van Kolck, Ann. Rev. Nucl. Part. Sci. 52, 339 (2002).
\bibitem{Epe06} E. Epelbaum, Prog. Part. Nucl. Phys. 57, 654 (2006).
\bibitem{Fin06} P. Finelli, N. Kaiser, D. Vretenar, W. Weise, Nucl. Phys. A 770, 1 (2006).
\bibitem{Epe09} E. Epelbaum, H.-W. Hammer, U.-G. Mei{\ss}ner, Rev. Mod. Phys. 81, 1773 (2009).
\bibitem{Cor09} L. Coraggio, A. Covello, A. Gargano, N. Itaco, T.T.S. Kuo, Prog. Part. Nucl. Phys. 62, 135 (2009).
\bibitem{Woo97} P.J. Wood, C.N. Davids, Ann. Rev. Nucl. Part. Sci. 47, 541 (1997).
\bibitem{Han03} P.G. Hansen and J. A. Tostevin, Annu. Rev. Nucl. Part. Sci. 53, 219 (2003).
\bibitem{Aum05} T. Aumann, Eur. Phys. J. A 26, 441 (2005).
\bibitem{Kee07} N. Keeley, R. Raabe, N. Alamanos, J.L. Sida, Prog. Part. Nucl. Phys. 59, 579 (2007).
\bibitem{Paa07} N. Paar, D. Vretenar, E. Khan, G. Colo, Rep. Prog. Phys. 70, 691 (2007).
\bibitem{Ben07} M.A. Bentley, S.M. Lenzi, Prog. Part. Nucl. Phys. 59, 497 (2007).
\bibitem{Bla08} B. Blank, M.J.G. Borge, Prog. Part. Nucl. Phys. 60, 403 (2008).
\bibitem{Gad08} A. Gade, Th. Glasmacher, Prog.Part. Nucl. Phys. 60, 161 (2008).
\bibitem{Kee09} N. Keeley, N. Alamanos, K.W. Kemper, K. Rusek, Prog. Part. Nucl. Phys. 63, 396 (2009).
\bibitem{Els34} W. Elsasser, J. de Phys. et Rad. 5, 625 (1934).
\bibitem{Goe48} M. Goeppert-Mayer, Phys. Rev. 74, 235 (1948).
\bibitem{Kae89} F. K\"appeler, H. Beer, K. Wisshak, Rep. Prog. Phys. 52, 945 (1989).
\bibitem{Gra07} H. Grawe, K. Langanke, G. Martinez-Pinedo, Rep. Prog. Phys. 70, 1525 (2007).
\bibitem{BBFH}  E.M. Burbidge, G. R. Burbidge, W. A. Fowler, and F. Hoyle, Rev. Mod. Phys. 29, 547 (1957).
\bibitem{Che95} B. Chen et al.,  Phys. Lett. B 355, 37 (1995).
 %J. Dobaczewski, K.-L. Kratz, K. Langanke, B. Pfeiffer, F.-K. Thielemann, P. Vogel,
\bibitem{Pfe97} B. Pfeiffer, K.-L. Kratz, F.-K. Thielemann, Z. Phys. A 357, 235 (1997).
\bibitem{Pfe01} B. Pfeiffer, et al., Nuclear Phys. A 693, 282 (2001).
\bibitem{Cow91} J.J. Cowan, F.-K. Thielemann, J.W. Truran, Phys. Rep. 208, 267 (1991).
\bibitem{Mey92} B.S. Meyer, G.J. Mathews, W.M. Howard, S.E. Woosley, R.D. Hoffman, Astrophys. J. 399, 656 (1992).
\bibitem{Woo92} S.E. Woosley, R.D. Hoffmann, Astrophys. J. 395, 202 (1992).
\bibitem{Pan08} I.V. Panov, H.-Th. Janka, Astronomy $\&$ Astrophysics 494, 829 (2009).
\bibitem{Sur08} R. Surman,  G.C. McLaughlin,  M. Ruffert,  H.-Th. Janka,   W.R.
Hix, Astrophyscial Journal Letters 679, L117 (2008).
\bibitem{Rin80} P. Ring, P. Schuck, The nuclear many-body problem, Springer-Verlag, New Work (1980), p.3.
\bibitem{Goe49} M. Goeppert-Mayer, Phys. Rev. 75, 1969 (1949).
\bibitem{Hax49} O. Haxel, J.H.D. Jensen, H.E. Suess, Phys. Rev. 75, 1766 (1949).
\bibitem{Ben03} M. Bender, P.H. Heenen, P.G. Reinhard, Rev. Mod. Phys. 75, 121 (2003).
\bibitem{Vre05} D. Vretenar, A.V. Afanasjev, G.A. Lalazissis, P. Ring, Phys. Rep. 409, 101 (2005).
\bibitem{Pie93} S.C. Pieper and V. R. Pandharipande, Phys. Rev. Lett. 70, 2541 (1993).
\bibitem{Ots09a} T. Otsuka, T. Suzuki, M. Honma, Y. Utsuno, N. Tsunoda, K. Tsukiyama, M. Hjorth-Jensen, arXiv:0908.2607 [nucl-th].
\bibitem{Wil84} B.H. Wildenthal, Prog. Part. Nucl. Phys. 11, 5 (1984).
\bibitem{Bru55} K.A. Brueckner, Phys. Rev. 97, 1353 (1955).
\bibitem{Gol57} J. Goldstone, Proc. Roy. Sot. A 293, 267 (1957).
\bibitem{Hjo95} M. Hjorth-Jensen, T.T.S. Kuo, E. Osnes, Physics Reports 261, 125 (1995).
\bibitem{Bro01} B.A. Brown, Prog. Part. Nucl. Phys. 47, 517 (2001).
\bibitem{Dea04} D.J. Dean, T. Engeland, M. Hjorth-Jensen, M.P. Kartamyshev, E. Osnes, Prog. Part. Nucl. Phys. 53, 419 (2004).
\bibitem{Cau05} E. Caurier, G. Martinez-Pinedo, F. Nowacki, A. Poves,  A.P. Zuker, Rev. Mod. Phys. 77, 427 (2005).
\bibitem{Bog01} S. Bogner, T.T.S. Kuo, L. Coraggio, Nucl. Phys. A 684, 432c (2001).
\bibitem{Bog02} S. Bogner, et al., Phys. Rev. C 65,  051301(R) (2002).
\bibitem{Bog03} S. Bogner, T.T.S. Kuo, A. Schwenk, Phys. Rep. 386, 1 (2003).
\bibitem{Hey87} K. Heyde, J. Jolie, J. Moreau, J. Ryckebusch, M. Waroquier, P. Van Duppen and, M. Huyse, J. L. Wood, Nucl. Phys. A466, 189 (1987).
\bibitem{Pov81} A. Poves, A. Zuker, Phys. Rep. 70, 235 (1981).
\bibitem{Uts99} Y. Utsuno, T. Otsuka, T. Mizusaki, M. Honma, Phys. Rev. C 60, 054315 (1999).
\bibitem{Ots01} T. Otsuka, R. Fujimoto, Y. Utsuno, B.A. Brown, M. Honma, T. Mizusaki, Phys. Rev. Lett. 87, 082502 (2001).
\bibitem{Ots05} T. Otsuka, T. Suzuki, R. Fujimoto, H. Grawe, Y. Akaishi, Phys. Rev. Lett. 95, 232502 (2005).
\bibitem{Ots06} T. Otsuka, T. Matsuo, D. Abe, Phys. Rev. Lett. 97, 162501 (2006).
\bibitem{Ots10} T. Otsuka et al.,  Phys. Rev. Lett. 104, 012501 (2010).
 %T. Suzuki, M. Honma, Y. Utsuno, N. Tsunoda, K. Tsukiyama, M. Hjorth-Jensen,
\bibitem{Bla06} K. Blaum, Physics Reports 425, 1 (2006).
\bibitem{Bla10} K. Blaum, Contemp. Phys. 51, 149 (2010).
\bibitem{Fra87} B. Franzke, Nucl. Instr. and Meth. B 24/25 (1987) 18.
\bibitem{Fra08} B. Franzke, H. Geissel, G. M\"unzenberg, Mass Spectr. Reviews 27, 428 (2008).
\bibitem{Hau00} M. Hausmann et al., Nucl. Instr. Meth. A 446, 569 (2000).
\bibitem{Rad00} T. Radon et al., Nucl. Phys. A 677, 75 (2000).
\bibitem{AME03} G. Audi, A.H. Wapstra and C. Thibault, Nucl. Phys. A 729, 337 (2003).
\bibitem{Ald56} K. Alder, A. Bohr, T. Huus, B. Mottelson, A. Winther, Rev. Modern Phys. 28, 432 (1956).
\bibitem{Ald66} K. Alder, A. Winther, Coulomb Excitation, Academic Press, New York, 1966.
\bibitem{Bro91} J.A. Brown et al., Phys. Rev. Lett. 66, 2452 (1991).
\bibitem{Osh92} M. Oshima et al., Nucl. Instr. Meth. A 312, 425 (1992).
\bibitem{Nie05} O. Niedermaier et al., Phys. Rev. Lett. 94, 172501 (2005).
\bibitem{Mot95} T. Motobayashi et al., Phys. Lett. B 346, 9 (1995).
\bibitem{Gla98} T. Glasmacher, Annu. Rev. Nucl. Part. Sci. 48, 1 (1998).
\bibitem{Ton78} F. Tondeur, Z. Phys. A 288, 97 (1978)
\bibitem{Hae89} P. Haensel and J. L. Zdunik, Astron. Astrophys. 222, 353 (1989).
\bibitem{Dob94} J. Dobaczewski, I. Hamamoto, W. Nazarewicz, and J. A. Sheikh, Phys. Rev. Lett. 72, 981 (1994).
\bibitem{Pea96} J.M. Pearson, R. C. Nayak, and S. Goriely, Phys. Lett. B 387, 455 (1996).
\bibitem{Lal98} G.A. Lalazissis, D. Vretenar, W. P\"oschl, and P. Ring, Phys. Lett. B 418, 7 (1998).
\bibitem{Sor08} O. Sorlin, M.-G. Porquet, Prog. Part. Nucl. Phys. 61, 602 (2008).
\bibitem{Oza00} A. Ozawa, T. Kobayashi, T. Suzuki, K. Yoshida, I. Tanihata, Phys. Rev. Lett. 84, 5493 (2000).
\bibitem{Hof09} C.R. Hoffman, et al. Phys. Lett. B 672, 17 (2009).
\bibitem{Kan09} R. Kanungo et al., Phys. Rev. Lett. 102, 152501 (2009).
\bibitem{Gei92} H. Geissel et al., Nucl. Instr. Meth. B 70, 286 (1992).
\bibitem{Jan09} R.V.F. Janssens, Nature 459, 1069 (2009).
\bibitem{War90} E. K. Warburton, J. A. Becker, and B. A. Brown, Phys. Rev. C 41, 1147 (1990).
\bibitem{Sch09} W. Schwerdtfeger et al., Phys. Rev. Lett. 103, 012501 (2009).
\bibitem{Doo09} P. Doornenbal et al., Phys. Rev. Lett. 103,  032501 (2009).
\bibitem{Kan10} R. Kanungo et al., Phys. Lett. B 685, 253 (2010).
\bibitem{Hon02} M. Honma, T. Otsuka, B.A. Brown, T. Mizusaki, Phys. Rev. C 65, 061301(R) (2002).
\bibitem{Pov01} A. Poves et al., Nucl. Phys. A 694, 157 (2001).
\bibitem{Cau02} E. Caurier et al., Eur. Phys. J. A 15, 145 (2002).
\bibitem{Pov05} A. Poves, F. Nowacki, E. Caurier, Phys. Rev. C 72, 047302 (2005).
\bibitem{Mai09} P. Maierbeck et al., Phys. Lett. B 675,  22 (2009).
\bibitem{Dil03} I. Dillmann, et al., Phys. Rev. Lett. 91, 162503 (2003).
\bibitem{Jun07} A. Jungclaus, et al., Phys. Rev. Lett. 99, 132501 (2007).
\bibitem{Sch04} J.P. Schiffer, et al., Phys. Rev. Lett. 92, 162501 (2004).
\bibitem{Hos05} P.T. Hosmer et al., Phys. Rev. Lett. 94, 112501 (2005).
\bibitem{Wal07} J. Van de Walle et al., Phys. Rev. Lett. 99, 142501 (2007).
\bibitem{Bar08} S. Baruah et al., Phys. Rev. Lett. 101, 262501 (2008).
\bibitem{Hak08} J. Hakala et al., Phys. Rev. Lett. 101, 052502 (2008).
\bibitem{Ste08} S.J. Steer et al., Phys. Rev. C 78, 061302 (2008).
\bibitem{Ald09} N. Al-Dahan et al., Phys. Rev. C 80, 061302 (2009).
\bibitem{Pod09} Z. Podolyak et al., Phys. Lett. B 672, 116 (2009).
\bibitem{Bla00} B. Blank et al., Phys. Rev. Lett. 84, 1116 (2000).
\bibitem{Dos05} C. Dossat et al., Phys. Rev. C 72, 054315 (2005).
\bibitem{Baz08} D.Bazin et al., Phys.Rev.Lett. 101, 252501 (2008)
\bibitem{Fae08} T. Faestermann et al., contribution to the 5th Int. Conf. on Exotic
 Nuclei and Atomic Masses (ENAM'08), Septmebter 7-13, 2008, Ryn, Poland.
\bibitem{Hin10} Ch. Hinke, Dissertation, TU M\"unchen, 2010.
\bibitem{Epp10} K. Eppinger, Dissertation, TU M\"unchen, 2010.
\bibitem{Sob07} A. Sobiczewski, K. Pomorski, Prog. Part. Nucl. Phys. 58, 292 (2007).
\bibitem{Hof00} S. Hofmann, G. M\"unzenberg, Rev. Mod. Phys. 72, 733 (2000).
\bibitem{Oga07} Yu. Oganessian, J. Phys. G 34, R165 (2007).

\end{thebibliography}
\end{document}